\documentclass[prd,twocolumn,floats,floatfix,showpacs,nofootinbib]{revtex4}
\usepackage{graphicx}
\usepackage{dcolumn}
\usepackage{bm}
\usepackage{graphics}
\usepackage{slashed}
\usepackage{amssymb}
\usepackage{natbib}
\usepackage{amsmath}
\usepackage{url}
\usepackage{color}
\newcommand{\bea}{\begin{eqnarray}}
\newcommand{\ena}{\end{eqnarray}}
\newcommand{\bean}{\begin{eqnarray*}}
\newcommand{\enan}{\end{eqnarray*}}

\begin{document}

\title{Beyond Analog Gravity: The Case of Exceptional Dynamics}

\author{M. Novello\footnote{M. Novello is Cesare Lattes ICRANet
Professor: novello@cbpf.br} and
E.Goulart\footnote{egoulart@cbpf.br}}

\affiliation{Instituto de Cosmologia Relatividade Astrofisica ICRA -
CBPF\\ Rua Dr. Xavier Sigaud, 150, CEP 22290-180, Rio de Janeiro,
Brazil}

\date{\today}

\begin{abstract}
We show that it is possible to go beyond the propagation aspects usually contemplated in the analog models of gravity.  Until this moment, the emergence of a metric appears in the description of excitations around a given background solution or in the study of field discontinuities in the geometrical optics regime. We now overcome some limitations of the above analogies and exhibit the form of a Lagrangian that describes the dynamics of a self-interacting field $
\varphi $ as an interaction between $ \varphi $ and its associated
effective metric $ \widehat{g}^{\mu\nu}.$ In other words the
non-linear equation of motion of the field may be interpreted as the
gravitational influence on $ \varphi $ by its own effective metric
which, in our scheme becomes an active partner of the dynamics of $
\varphi. $

 \end{abstract}

 \maketitle

\section{Introduction}
In recent years, intense activity on analogue gravity has been
developed \cite{AG}.This was concentrated in nonlinear
electrodynamics, acoustics, hydrodynamics, optics inside media and
various condensed matter systems. This term (analog gravity) implies
 the description of distinct physical processes as modifications of
the metrical structure of the background space-time. Until now this
analogy had been limited to perturbative aspects restricting it to
the propagation of excitations (photons and quasi-particles) through a medium or a background field configuration. In
this paper we give the first example which goes beyond this
limitation describing dynamical features of fields in terms of their respective effective
metric\footnote{We would like to stress that this dynamical aspect we are investigating has nothing to do with mimicking Einstein's equations through effective metrics.}. The main steps to achieve this result are the following:
\begin{itemize}
\item{Consider a non linear field theory described in a flat
Minkowski background;}
\item{Note that the propagation of the discontinuities of the field
yields the irruption of a second metric $ \widehat{g}^{\mu\nu}$ such
that the path of the waves are null geodesics in this effective
geometry;}
\item{There exists a special class of theories such that its
corresponding dynamics (which we will deserve the name exceptional)
can be described alternatively as the gravitational interaction of
the field in a given curved geometry;}
\item{The geometry of such gravitational space-time is precisely the effective
metric $ \widehat{g}^{\mu\nu}.$ }
\end{itemize}
Thus the claim of the present paper is that the self-interaction
described by exceptional dynamics is described in an equivalent way
as the gravitational interaction of the field with its own effective
metric.

For pedagogical reasons we limit our analysis here to the case of
non linear scalar fields. Generalization to the case of other
fields, like non-linear Electrodynamics will be described elsewhere.

\section{The kinematical analogy}

 Let us give a simple example. Consider a scalar field $
\varphi $ propagating in a flat spacetime whose dynamics is provided by the non-linear Lagrangian
\cite{noncanonicalSUGRA},


$$ L = L(w) $$
where $$ w \equiv \partial_{\mu} \varphi \, \partial_{\nu} \varphi
\, \eta^{\mu\nu} $$
is the canonical kinetical term. The equation of motion for $\varphi$ reads:

\begin{equation}\label{EqMotion}
\partial_\mu \Bigl( L_{w} \, \partial_\nu
\varphi \, \eta^{\mu\nu} \Bigr) = 0
\end{equation}
where $L_w$ denotes the first derivative of $L$ with respect to $w.$ By derivating the left hand side we obtain the explicit form
\begin{equation}\label{eom}
L_{w}\square\varphi+2L_{ww}\partial^{\mu}\varphi\partial^{\nu}\varphi\partial_{\mu}\partial_{\nu}\varphi=0,
\end{equation}
with $\square\equiv\eta^{\mu\nu}\partial_{\mu}\partial_{\nu}$. This
constitutes a quasi-linear second order partial differential
equation for $\varphi$. We are interested in evaluating the
characteristic surfaces of wave propagation in this theory. The most
direct and elegant way to pursue this goal is to use the Hadamard
formalism \cite{Hadamard}.


 Let $\Sigma$ be a surface of discontinuity of
the scalar field $\varphi.$ The discontinuity of an arbitrary
function $f$ is given by:
\begin{equation}\label{DiscDef}
\left[ f(x) \right]_\Sigma = \lim_{\epsilon \to 0^+} \bigl( f(x +
\epsilon) - f(x - \epsilon)\bigr)
\end{equation}
The field $\varphi$ and its first derivative $\partial_\mu \varphi$
are continuous across $\Sigma$, while the second derivative presents
a discontinuity:
\begin{eqnarray}\label{CondHadamard}
\left[ \varphi \right]_\Sigma &=& 0,\\
\left[ \partial_\mu \varphi \right]_\Sigma &=& 0,\\
\left[ \partial_\mu\partial_\nu\varphi \right]_\Sigma &=& k_\mu
k_\nu \xi(x),
\end{eqnarray}
where $ k_\mu := \partial_\mu \Sigma$ is the propagation vector and
$\xi(x)$ the amplitude of the discontinuity. From the above
conditions we obtain that both $L_{w}$ and $L_{ww}$ are continuous
functions across $\Sigma$. Using these discontinuity properties in
the equation of motion (\ref{eom}) it follows that only the second
order derivative terms contribute. We obtain the relation
\begin{equation}
L_{w}\eta^{\mu\nu}\left[ \partial_\mu\partial_\nu\varphi
\right]_\Sigma+ 2 \, L_{ww}\, \partial^\mu \varphi
\partial^\nu\varphi \, \left[
\partial_\mu\partial_\nu\varphi \right]_\Sigma=0.
\end{equation}
Thus, using Hadamard conditions it follows
\begin{equation*}
k_\mu k_\nu \Bigl(L_w \eta^{\mu\nu} + 2 L_{ww}
\partial^\mu\varphi\partial^\nu\varphi\Bigr) = 0,
\end{equation*}
This equation suggests the introduction of the effective metric
defined by:
\begin{equation}\label{gEff}
\widehat{g}^{\mu\nu} := L_w \eta^{\mu\nu} + 2 L_{ww}
\partial^\mu\varphi\partial^\nu\varphi
\end{equation}
Thus there are two distinct metrics in this framework: the
Minkowskian $ \eta^{\mu\nu} $ that enters in the dynamics of the
field $\varphi$  and the effective metric $ \widehat{g}^{\mu\nu} $
that controls the propagation of the waves. Note that, once the
vector of discontinuity $ k_{\mu}$ is a gradient, discontinuities of
the field $\varphi$ propagate through null geodesics in the
effective metric $\widehat{g}_{\mu\nu},$ i.e.
\begin{equation}
\hat{g}^{\mu\nu}k_{\alpha ;\mu}k_{\nu}=0,
\end{equation}
where ``;" stands for the covariant derivative evaluated with the
effective metric. The inverse $\widehat{g}_{\mu\nu}$ of (\ref{gEff})
is obtained through the condition
$\hat{g}^{\mu\alpha}\hat{g}_{\alpha\nu} = \delta^\mu_\nu:$
 \begin{equation}\label{gEffInv}
\widehat{g}_{\mu\nu} = \frac{1}{L_w} \, \eta_{\mu\nu} - \frac{2
L_{ww}}{L_{w} \,\Psi}
\partial_\mu\varphi\partial_\nu\varphi ,
\end{equation}
where we defined $\Psi := L_w + 2 w L_{ww}$. For latter reference we
note that the determinant of a mixed tensor $\textbf{T}=T^{\alpha}_{\phantom a\beta}$ may be expressed in terms of traces of its powers in the form
\begin{eqnarray}\label{determinant}
 - 4 \, det \textbf{T} &=& Tr (\textbf{T}^{4}) - \frac{4}{3} \,
Tr (\textbf{T}) \, Tr (\textbf{T}^{3}) \nonumber \\ &-& \frac{1}{2} \, \left(Tr
(\textbf{T}^{2})\right)^{2} + \left(Tr (\textbf{T})\right)^{2} \, Tr (\textbf{T}^{2})\\ &-&
\frac{1}{6} \, \left( Tr (\textbf{T}) \right)^{4}.
\end{eqnarray}
This is an immediate consequence of the Cayley-Hamilton theorem.
Applying this formula to the effective metric (\ref{gEff}) one
obtains, after a straightforward calculation
\begin{equation}\label{det}
\sqrt{ -\widehat{g}} = L_{w}^{-2} \, (1 + \beta \, w)^{-1/2},
\end{equation}
where we have defined $ \beta = 2 \, L_{ww} /L_{w} .$ Note that, the
square-root of the determinant is real only if the condition
\begin{equation}
1+2w \, L_{ww} /L_{w}>0
\end{equation}
is satisfied. This is the same as to guarantee the hiperbolicity of
the equations of motion and henceforth the existence of waves.
Nevertheless, we remark that the effective metric that controls the
propagation of these waves is not unique and is determined up to a
conformal factor. However, as it occurs in typical non-linear
theory, the dynamics is not conformal-invariant.

\section{Exceptional dynamics}

Let us note that the discontinuities of the field propagate in a
curved space-time, although the field $\varphi $ is described by a
non-linear theory in Minkowski geometry \cite{Novello2}.


In the framework of general relativity it is the presence of gravity
that allows the existence of curvature in the geometry. This has led
to the interpretation of the dispersion relation of non-linear
fields in terms of an effective metric as nothing but the
simplification of its description, that is, a matter of language.

 At this point we face the following question:
 is it be possible that among all non-linear theories one can select a
 special class such that the dynamics of the field itself is
 described in terms of the effective metric? That is, is it be
 possible to unify the dynamics of the field with the propagation of
 its waves such that just one metric appeared?\footnote{Note that this is not equivalent
 to the known property --- that occurs in hydrodynamics (and also in field theory) ---
 that the characteristic propagation of the field coincides with the characteristic propagation of its
 perturbations (see appendix for a detailed discussion).}

 If this is possible, then we are in the presence of a field that mimics a kind of gravitational interaction (coupling) by means of its own equation of motion. In other
 words, the equation of evolution of $ \varphi $ is equivalent to
 the gravitational interaction between $ \varphi $ and its effective metric.
 This means that the equation of motion
 (\ref{EqMotion}) can be written under the equivalent form
\begin{equation}\label{1fevereiro2011}
\widehat{\Box}\, \varphi \equiv     \frac{1}{\sqrt{-\hat{g}}} \,
\partial_\mu\Bigl(\sqrt{-\hat{g}} \,
  \partial_\nu \varphi \, \hat{g}^{\mu\nu}\Bigr) = 0.
\end{equation}
In general it is not possible to re-write the nonlinear equation
(\ref{EqMotion}) in the above form for an arbitrary lagrangian.
However, we will show next that there exist some special situations
where this implementation becomes feasible. We first note the
following relation
\begin{equation}
 \partial_{\nu}\varphi\hat{g}^{\mu\nu}=(L_{w}+2wL_{ww}) \partial_{\nu}\varphi\eta^{\mu\nu}
\end{equation}
It then follows that the dynamics described by (\ref{EqMotion}) and
 (\ref{1fevereiro2011}) will be the same, provided the Lagrangian satisfies the condition
\begin{equation}
L_{w}=\sqrt{-\hat{g}}(L_{w}+2wL_{ww}).
\end{equation}
Using the expression for the determinant (\ref{det}) the equivalence
is provided by the nonlinear differential equation for the
lagrangian
 \begin{equation}
 2 \, w \, L_{ww} + L_{w} - L_{w}^{5} = 0.
 \label{1fevereiro2}
 \end{equation}
We will call the system described by Lagrangians that satisfies
condition (\ref{1fevereiro2}) as Exceptional Dynamics. In other
words, non-linear systems described by exceptional dynamics may be
alternatively interpreted as fields gravitationally coupled to its
own effective geometry. Thus the field and the corresponding waves
agree in the interpretation that the geometry of the space-time is
given by $\widehat{g}_{\mu\nu}.$

Equation (\ref{1fevereiro2}) is such that the first derivative of
the lagrangian with respect to $w$  may be obtained explicitly.
Indeed, one obtains that
\begin{equation}\label{first}
L_{w}=\pm\frac{1}{(1-\lambda w^{2})^{\frac{1}{4}},
}
\end{equation}
where $\lambda$ is an arbitrary real positive constant. The general
solution of equation (\ref{1fevereiro2}) may be obtained as an
infinite series given by the hypergeometric function

\begin{equation}\label{excep}
L(w)=\pm wHyp2F1\left[\frac{1}{2},\frac{1}{4},\frac{3}{2},\lambda w^{2}\right].
\end{equation}
 We note that this dynamics is such that the admissible values of
  $ w $ are restricted to the domain $w^{2}<\lambda$. Thus, the theory
naturally avoids arbitrarily large values of the kinematical term.
Second, the lagrangian is a monotonic function of $ w $ as an
immediate consequence of equation $(\ref{first})$. Third, the theory
admits the linear theory as a limiting case for small values of $w$,
as we will soon see. Lastly, the resulting function is odd, i.e
$L(w)=L(-w)$.

A comparison between the exceptional lagrangian (\ref{excep}) and
the canonical linear lagrangian $L=w$ is given in figure I. From the
plot we see that, in fact, the new theory is described by a simple
function that possess many desirable features. We filled the space
between the functions to give a concrete idea of how they differ
from each other in this domain. Note that, although the maximum
difference happens in the boundaries $w^{2}=\lambda$, the
exceptional lagrangian is very similar to the linear one for small
values of $``w"$.

It is instructive to expand the exceptional lagrangian to obtain a
clear idea of its behavior in terms of the parameter $\lambda$. We
obtain
\begin{equation}\label{expansion}
L(w)=w+\frac{\lambda}{12}w^{3}+\frac{\lambda^{2}}{32}w^{5}+\frac{15\lambda^{3}}{896}w^{7}+O[x]^{9}
\end{equation}
Note that for small values of the constant, $\lambda<< 1 $ the
exceptional dynamics reduces to a cubic lagrangian\footnote{We could
arrive at this result assuming from the beginning a lagrangian in
the vicinity of the linear theory of the form $L(w)=w+\epsilon f(w)$
with $\epsilon^{2}<<\epsilon$ and solving the simplified equation
$wf_{ww}-2f_{w}=0$.}. The linear theory is obviously recovered when
$\lambda=0$, implying that both excitations (discontinuities) and
field dynamics are described by the same effective metric i.e. $\eta^{\mu\nu}$.\\

\begin{figure}\label{graph}
\center
\includegraphics[width=8cm,height=6cm]{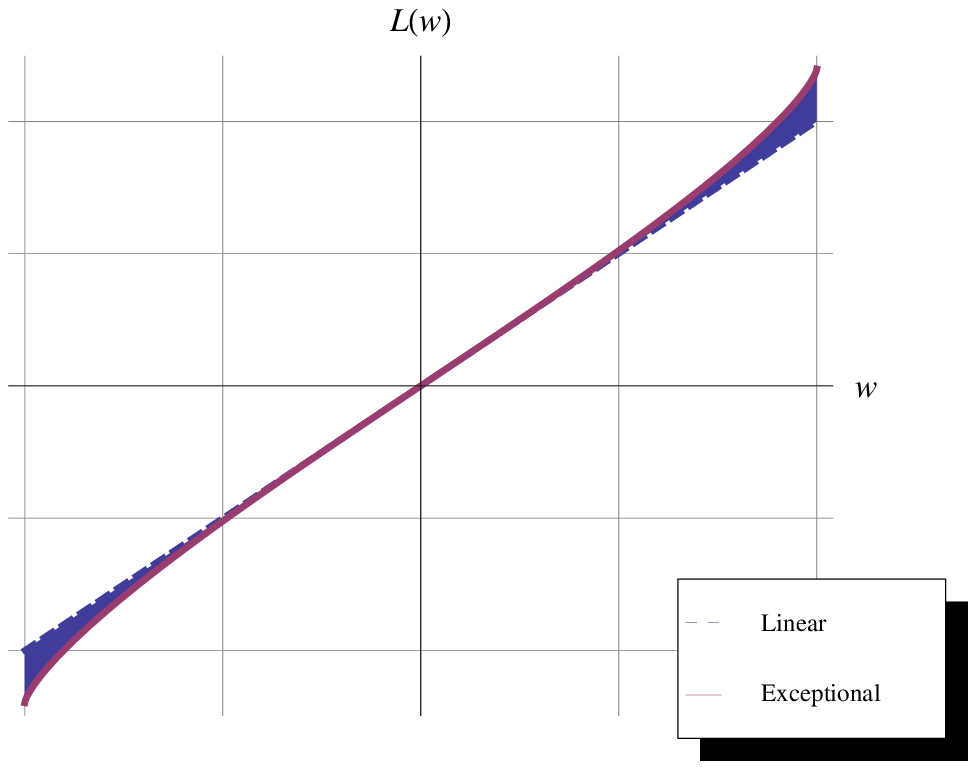}
\caption{\label{st}} Comparison between linear and exceptional
lagrangians.
\end{figure}

\section{Another simple example}

In the precedent section we examined the case in which a
self-interacting scalar field generates a geometrical arena for its
own propagation. The presence of an effective metric that controls
the propagation of the discontinuities of the field can be extended
in such a way that the original theory, describing the dynamics of
the scalar field in a flat Minkowski arena can be alternatively
described as the gravitational interaction of the field. Then we
have shown a non-expected result: the metric field that describes
this gravitational effect is nothing but the same effective metric
that controls the wave propagation.

In other words, our main achievement was to produce a scenario where
both the propagation of the field discontinuities and the field
dynamics are controlled by the same metric structure. The
nonlinearities are such that the equations in Minkowski space-time
mimic the dynamics of a ``free-field" embedded in a curved
space-time generated by the field itself.

Let us now provide another simple example of a physical system where
the exceptional dynamics may be relevant to real practical
situations and may be used to investigate dynamical aspects of
fields in curved space-times. The basic system consists of Maxwell's
electrodynamics inside (linear) dielectric medium in motion. The
quantities we will deal with consists in tensors $F_{\mu\nu}$ and
$P_{\mu\nu}$ and a normalized time-like vector field $v^{\mu}(x)$
that is $\eta_{\mu\nu}v^{\mu}v^{\nu}=1$ which represents the
velocity field of the material. We have

\begin{eqnarray*}
F_{\mu\nu} &=& E_{\mu}v_{\nu} - E_{\nu}v_{\mu} +\eta_{\mu\nu}{}^{\alpha\beta}v_{\alpha}B_{\beta}, \label{1}
\\
P_{\mu\nu} &=& D_{\mu}v_{\nu} - D_{\nu}v_{\mu} +\eta_{\mu\nu}{}^{\alpha\beta}v_{\alpha}H_{\beta} . \label{2}
\end{eqnarray*}
We restrict our analysis to the case of isotropic constitutive laws provided by
\begin{equation}
D_{\alpha}=\epsilon(x)E_{\alpha}\quad\quad B_{\alpha}=\mu(x)H_{\alpha}.
\end{equation}
The equations of motion are
\begin{equation}\label{max}
P^{\mu\nu}_{\phantom a\phantom a ,\nu}=0\quad\quad F_{[\mu\nu,\alpha]}=0,
\end{equation}
and the study of field discontinuities inside such material leads to Gordon's metric
\begin{eqnarray*}
\hat{g}^{\mu\nu}&=&\eta^{\mu\nu}+(\mu\epsilon-1)v^{\mu}v^{\nu}.
\end{eqnarray*}
Thus, the wave fronts inside an isotropic, heterogeneous and linear
material is described by a null vector $k_{\mu}$ with respect to the
effective metric i.e. $\hat{g}^{\mu\nu}k_{\mu}k_{\nu}=0$.
Furthermore, it is immediate to show that $k_{\mu}$ is a geodesic
with respect to $\hat{g}_{\mu\nu}$. This result is well known and
was obtained several times in the literature of analog models,
enabling the study of kinematical aspects of fields in the presence
of gravitation. On the other hand, there exist various situations,
typical of certain class of materials, where it is possible to go
beyond this kinematical analogy. In these cases the field dynamics
itself is described by its dependence on the effective metric, as in
the case of exceptional dynamics.

In fact, using the determinant formula (\ref{determinant}) it follows that
\begin{equation}\label{D}
\hat{g}\equiv det(\hat{g}_{\mu\nu})=-\frac{1}{\mu\epsilon}.
\end{equation}
Also, from the definition of the effective metric we obtain the identity
\begin{equation}\label{P}
\hat{F}^{\mu\nu}\equiv \hat{g}^{\mu\alpha}\hat{g}^{\nu\beta}F_{\alpha\beta}=\mu P^{\mu\nu}.
\end{equation}
Using (\ref{D}) and (\ref{P}) simultaneously it is possible to write
the first of Maxwell equations (\ref{max}) in the form
\begin{equation}
\left(\frac{1}{\mu} \, \hat{F}^{\mu\nu}\right)_{,\nu}=0.
\end{equation}
In the case of impedance matched materials where the ratio
$\epsilon/\mu$ is constant, we can re-write this equation in the
very suggestive form
\begin{equation}
(\sqrt{-\hat{g}} \, \hat{F}^{\mu\nu})_{,\nu}=0,
\end{equation}\\
Finally the complete set of Maxwell equations (\ref{max}) in this
medium can be written as
\begin{equation}
\hat{\nabla}_{\nu}\hat{F}^{\mu\nu}=0\quad\quad \hat{\nabla}_{[\alpha}\hat{F}_{\mu\nu]}=0,
\end{equation}
where $\hat{\nabla}_{\alpha}$ is the covariant derivative written in
terms of $\hat{g}_{\mu\nu}$. This means that the effective metric,
that describes the characteristics, has an active part in the very
description of the field dynamics. This situation can be used as a
tool to investigate simultaneously kinematical and dynamical aspects
of fields interacting with gravity in laboratories.

\section{Conclusion}

Let us summarize the novelty of our analysis in a broad sense:
\begin{itemize}
\item{For any field theory described on a Minkowski background by a non-linear Lagrangian $ L = L(w)$
 the discontinuity of $\varphi$ propagates as null geodesics in an effective metric $
\widehat{g}^{\mu\nu};$}
\item{As such, the theory presents a duplicity of metrics: the field is described in flat
Minkowski space-time and its corresponding waves propagate as null
geodesics in a curved geometry;}
\item{It is possible to unify the description of the dynamics of $ \varphi$ in such a way
that only one metric appears. This is possible for those Lagrangians
that represent exceptional dynamics;}
\item{In this case, the self-interaction of $ \varphi$ is described
equivalently as if it were interacting minimally with its own
effective geometry, allowing the interpretation in terms of an
emergent gravitational phenomenon.}
 \end{itemize}

The structure of the nonlinear equations of motion suggests that the
previous procedure can be adapted to other structures like non
linear electrodynamics in a moving dielectric, vector and tensor
field theories. We will develop this idea in another paper.

\section{Appendix}
In this appendix we would like to emphasize that the possibility to
describe the dynamics of scalar field as an interaction with its own
effective metric --- and as such, mimic gravitational interaction --
is not valid for all types of non-linear theory but only for a
specific class -- which we called exceptional dynamics.

On the other hand, one should not confuse such property with a
generic one concerning the propagation of the discontinuities.
Indeed, the condition of a given Lagrangian to describe an
exceptional dynamics is completely independent of the known fact
that \emph{the characteristic propagation of the field is similar to
the characteristic propagation of its (linear) perturbations. That
is both discontinuities are guided by the same effective metric
$\hat{g}_{\mu\nu}$.} Let us prove this.

The characteristic surfaces of the nonlinear equation of motion
(\ref{EqMotion}) are obtained by the analysis of its
discontinuities. As we shown in a precedent section these surfaces
are such that
\begin{equation}
\hat{g}^{\mu\nu}k_{\mu}k_{\nu}=0.
\end{equation}
with $\hat{g}^{\mu\nu}$ given by (\ref{gEff}). We now turn to a
diferent question. Let us consider a perturbation of the field
$\varphi$, i.e.
\begin{equation}
\varphi\rightarrow \varphi+\delta\varphi.
\end{equation}
If we linearize equation  (\ref{EqMotion}) around a given background
solution $\varphi$, retaining only first order terms in
$\delta\varphi$, then what are the characteristic surfaces for the
perturbation $ \delta\varphi?$ The linearized equations reads
\begin{eqnarray}
&&(L_{w}\eta^{\mu\nu}+2L_{ww}\varphi^{,\mu}\varphi^{,\nu})\delta\varphi_{,\mu,\nu}+\\
+&&2L_{ww}(\square\varphi\varphi^{,\mu}+2\varphi^{,\mu}_{\phantom a ,\nu}\varphi^{,\nu})\delta\varphi_{,\mu}+\\
+&&4L_{www}\varphi_{,\alpha ,\nu}\varphi^{,\alpha}\varphi^{,\nu}\varphi^{,\mu}\delta\varphi_{,\mu}=0
\end{eqnarray}
Applying Hadamard's analysis to this new equation i.e.
\begin{eqnarray}\label{Hadamard}
\left[ \delta\varphi \right]_\Sigma &=& 0,\\
\left[ \partial_\mu \delta\varphi \right]_\Sigma &=& 0,\\
\left[ \partial_\mu\partial_\nu\delta\varphi \right]_\Sigma &=& k_\mu
k_\nu \xi(x),
\end{eqnarray}
we obtain the well-known result that the perturbations
$\delta\varphi$ are governed by the same effective metric. Thus, the
characteristic propagation of the field is equivalent to the
characteristic propagation of its own disturbances. This fact is
valid for any theory and is totally independent of the specification
of the lagrangian.

\section{acknowledgements}
MN would like to thank FINEP, CNPq and FAPERJ and EG FAPERJ for
their financial support.

\end{document}